\documentclass[sigconf,nonacm]{acmart}

\usepackage{amsmath,amssymb,amsfonts}
\usepackage{graphicx}
\usepackage{textcomp}
\usepackage{xcolor}
\usepackage{url}
\usepackage{booktabs}
\usepackage{multirow}
\usepackage{subfigure, caption}
\usepackage{array}

\usepackage{pifont}  
\usepackage[most]{tcolorbox}
\usepackage{balance}
\usepackage{textcomp}
\usepackage{comment} 
\usepackage{float}
\usepackage{algpseudocode}
\usepackage{rotating}
\usepackage{wasysym}
\usepackage{booktabs}
\usepackage{hyperref}
\usepackage{subcaption}
\usepackage{afterpage}
\usepackage{enumitem}
\usepackage{stfloats}
\usepackage{comment} 
\usepackage{rotating}
\usepackage{appendix}
\usepackage{amssymb}
\usepackage{rotating}

\AtBeginDocument{%
  \providecommand\BibTeX{{%
    \normalfont B\kern-0.5em{\scshape i\kern-0.25em b}\kern-0.8em\TeX}}}
    
\begin{document}

\title{Unveiling Usability Challenges in Web Privacy Controls}

\author{Rahat Masood}
\affiliation{%
  \institution{The University of New South Wales}
  \city{Sydney}
  \state{NSW}
  \country{Australia}
}
\email{rahat.masood@unsw.edu.au}

\author{Sunday Oyinlola Ogundoyin}
\affiliation{%
  \institution{Macquarie University}
  \city{Sydney}
  \state{NSW}
  \country{Australia}
}
\email{sunday.ogundoyin@hdr.mq.edu.au}

\author{Muhammad Ikram}
\affiliation{%
  \institution{Macquarie University}
  \city{Sydney}
  \state{NSW}
  \country{Australia}
}
\email{muhammad.ikram@mq.edu.au}

\author{Alex Ye}
\affiliation{%
  \institution{The University of New South Wales}
  \city{Sydney}
  \state{NSW}
  \country{Australia}
}
\email{alex.ye@unsw.edu.au}

\begin{abstract}
With the increasing concerns around privacy and the enforcement of data privacy laws, many
websites now provide users with privacy controls. However, locating these controls can be challenging, as
they are frequently hidden within multiple settings and layers. Moreover, the lack of standardization means
these controls can vary widely across services. The technical or confusing terminology used to describe
these controls further complicates users' ability to understand and use them effectively. 

This paper presents a large-scale empirical analysis investigating usability challenges of web privacy controls across 18,628 websites. While aiming for a multi-scenario view, our automated data collection faced significant hurdles, particularly in simulating sign-up and authenticated user visits, leading to more focused insights on guest visit scenarios and challenges in automated capture of dynamic user interactions. Our heuristic evaluation of three different user visit scenarios identifies significant website usability issues. Our results show that privacy policies are most common across all visit scenarios, with nudges and notices being prevalent in sign-up situations. We recommend designing privacy controls that: enhance awareness through pop-up nudges and notices; offer a table of contents as navigational aids and customized settings links in policies for more informed choice; and ensure accessibility via direct links to privacy settings from nudges.
\end{abstract}


\maketitle
\section{Introduction}
\label{sec:intro}
With the widespread proliferation of online platforms and the increasing volume of personal data shared by users, service providers and developers are increasingly focused on designing robust mechanisms to safeguard end-user privacy. In the domain of online privacy, multiple stakeholders and concerns intersect. Users express significant concern regarding the safety and handling of their personal information~\cite{pewresearch}, while regulatory frameworks—such as the General Data Protection Regulation (GDPR)~\cite{gdpr}—impose stringent compliance requirements on digital services, including the need for explicit and informed user consent for data processing. Developers, in turn, strive to satisfy these regulatory obligations, but often encounter difficulties in implementing privacy-preserving mechanisms that are both effective and user-friendly~\cite{ataei2018complying}.

A growing body of research has examined the usability of online privacy controls, addressing features such as data deletion, opt-out mechanisms, and cookie consent interfaces. For example, the empirical study by Habib et al.\cite{habib2019empirical} highlights issues such as inconsistent interface formats and obfuscated links, which complicate users’ ability to exercise informed privacy choices. Even platforms employing Consent Management Providers (CMPs) to streamline cookie preference selection have been found to log user consent inaccurately, even when users have explicitly opted out~\cite{matte2020cookie}. Additional usability challenges include the low readability of privacy policies~\cite{ali2020readability}, poorly designed cookie opt-out interfaces~\cite{feng2021design}, and the presence of deceptive design patterns (i.e., \textit{dark patterns}) that subtly manipulate users into relinquishing privacy protections~\cite{rinehart2022bringing, toth2022dark}.

Despite these contributions, existing literature tends to focus on isolated privacy controls, thereby overlooking the holistic user experience across multiple privacy dimensions. This narrow focus limits our understanding of the comprehensive effectiveness of privacy controls on websites. As a result, a site may perform well in one aspect—such as providing a readable privacy policy—while falling short in others, such as accessibility or navigability of privacy settings. For instance, although \texttt{fitbit.com}, a prominent health and fitness platform, offers a clearly structured and comprehensible privacy policy, our analysis reveals that accessing its privacy settings requires navigating through five separate clicks, posing a potential barrier to usability.

This study addresses critical gaps in understanding the usability and implementation of privacy controls on websites. Specifically, we conduct a large-scale empirical analysis focusing on four primary types of privacy controls: privacy nudges, privacy policies, privacy notices, and privacy settings. Our analysis encompasses 18,628 websites, drawn from widely used and diverse popularity rankings, including Alexa\cite{alexa1m}, Hispar\cite{hispar}, and Majestic Million~\cite{majestic1m}. Through this extensive investigation, we assess both the presence and usability of these controls across various website categories and user interaction contexts.

To enable this assessment, we developed three data collection templates that simulate distinct user visit scenarios: \textit{(i)} guest visits (where users browse without logging in), \textit{(ii)} account sign-up processes, and \textit{(iii)} authenticated sessions (where users are logged in). Based on the data gathered via these templates, we applied a custom usability evaluation framework, informed by prior methodologies~\cite{albesher2021evaluating, redmiles2020comprehensive}, to systematically assess the design and accessibility of privacy controls. While our data collection methodology aimed to cover all three user visit scenarios across this entire dataset, practical challenges, particularly in automating account creation and maintaining authenticated sessions, led to varying effective sample sizes for each scenario, as detailed in Section~\ref{sec:dcmethod}. Consequently, the broadest analysis of privacy controls applies primarily to the guest visit scenario, and it was not possible to provide categorical insights for the registered user scenario due to the inherent difficulty in consistently obtaining data for this complex interaction phase across the full dataset.

The following research questions drive our investigation:

\begin{itemize}
\item $\mathbf{RQ1}.$ \textit{To what extent does a website’s ranking influence the usability and implementation of its privacy controls?}
\item $\mathbf{RQ2}.$ \textit{How does the usability of privacy controls vary across different website categories?}
\item $\mathbf{RQ3}.$ \textit{What is the presence and usability of privacy controls across different user visit scenarios?}
\end{itemize}

To answer these questions, we employed automated web crawling techniques to extract structured data from a large and diverse set of web platforms. Our analysis reveals notable patterns in the implementation and accessibility of privacy controls. However, while our coverage of guest visit scenarios is comprehensive, automated crawling for more dynamic interactions—particularly the sign-up and authenticated user flows—posed technical challenges. For example, the successful capture rate for sign-up scenarios was limited to 20.3\%, primarily due to anti-bot mechanisms and dynamic content rendering. As a result, while findings from guest visit scenarios are robust and broadly representative, insights derived from sign-up and logged-in user states are based on smaller and potentially less generalizable subsets of data. 

Our primary contributions are as follows:
\begin{itemize}[left=0pt]
 \item \textbf{A comprehensive, large-scale usability analysis.} We conduct an unprecedented empirical analysis of privacy controls across 18,628 websites, spanning diverse categories and user interaction scenarios (guest visits, sign-up processes, and authenticated sessions). Our study is unique in taking a holistic approach, evaluating four distinct privacy controls—nudges, notices, policies, and settings—against five key usability principles: Awareness, Efficiency, Comprehension, Functionality, and Choice. This comprehensive methodology addresses critical gaps in existing literature that often focus on single privacy controls or limited scopes.
 \item \textbf{Unveiling pervasive challenges in user awareness and efficiency.} We empirically demonstrate significant usability flaws hindering users' ability to discover and access privacy controls. Privacy nudges, designed to enhance awareness, are strikingly scarce, present on only 5.8\% of top-ranked websites and a mere 2.6\% of guest visits for notices. This widespread absence means many users are simply unaware of privacy prompts. Even when controls exist, accessing them demands excessive user effort: a staggering 83.4\% of top-ranked websites require two clicks to access a privacy policy, and a concerning 42.3\% of websites fail to provide any direct links to privacy settings. This creates \textit{obfuscated links} and \textit{confusing navigation}, directly impeding user efficiency and increasing frustration, effectively hiding controls from active engagement.
 \item \textbf{Exposing critical gaps in user comprehension.} Our findings reveal a systemic failure in supporting user understanding of privacy information, directly undermining informed consent. Multilingual support is severely lacking, with only 22\% of top-ranked websites offering privacy policies in various languages, a figure that plummets to 8.4\% for low-ranked sites. Similarly, 78.4\% of websites do not offer settings in multiple languages. This forces non-proficient speakers to \textit{manually translate content}, severely limiting their ability to understand privacy terms. Navigational aids are virtually non-existent: a mere 5.5\% of top websites include a table of contents in their privacy policies, hindering efficient navigation through lengthy documents and requiring users to \textit{scroll through the entire privacy policy}. Furthermore, 58.5\% of websites fail to provide clear guidance on how to use privacy settings, directly undermining their effectiveness for users. 
 \item \textbf{Highlighting restrictions on user choice and autonomy.} We reveal how existing implementations limit meaningful user control over personal data, akin to \textit{dark patterns that manipulate privacy choices}. A concerning proportion of privacy nudges, notably 42.5\% on low-ranked websites, 70.6\% in Adult Content, and two-thirds in Education categories, restrict users to only an `Accept' option. This \textit{may restrict user autonomy} by forcing acceptance without alternatives. The pervasive absence of direct links to privacy settings (present on only 7.3\% of top websites and 3.4\% of low-ranked websites) underscores a significant barrier to users exercising granular control over their data.
 \end{itemize}

Our detailed analysis reveals how usability varies significantly by website ranking and category, from more accessible settings on Shopping and Government sites to more limited options on Education and Technology platforms. These findings offer invaluable advocacy for improvement in privacy control design and implementation across all website tiers and categories. We emphasize the crucial need for standardized, transparent, and user-friendly privacy practices across the web to empower users with meaningful choices and greater control over their data, aligning with global data protection regulations like GDPR. 

\section{Related work}
\label{sec:rwork}
Several studies have indeed examined the use cases and effectiveness of various privacy controls, shedding light on these usability challenges. 

\textit{Privacy nudges and notices.} Ioannou et al.~\cite{ioannou2021privacy} conducted a meta-analytic review of 78 papers to explore the efficacy of privacy nudges, which are important for understanding how nudges influence user behavior regarding personal information disclosure. In contrast, Adjerid et al.~\cite{AdjeridPrivacyNotice} conducted a series of experiments to analyze the role of privacy notices in assisting consumers with disclosure-related choices, such as sharing demographic information. Mehrnezhad et al.~\cite{mehrnezhad2020cross} analyzed privacy notices and tracking behaviors on browser and mobile app platforms, demonstrating that most web pages present the privacy consent banner inconsistently. Such inconsistency negatively impacts user comprehension and efficiency. 

\textit{Usability of opt-out, data deletion, and cookie management.} Researchers have also developed user-driven survey templates to assess the usability of privacy choices and privacy advice pieces. For example, when examining privacy choices, Hana et al.~\cite{habib2020s} utilized the cognitive and physical processes required to use privacy choices, highlighting issues with the usability of websites' opt-out and data deletion choices. Their earlier study (also cited as Habib et al.~\cite{habib2019empirical}) further revealed issues of inconsistent formats and obfuscated links that make it difficult and confusing for users to navigate privacy choices. 
Even websites utilizing Consent Management Providers (CMPs) to facilitate cookie preference selection can exhibit incorrect logging consent, even when users have opted out. Matte et al. found that 54\% of cookie banner implementations had at least one suspected violation, including pre-selected choices or the absence of an option to refuse consent, which directly impacts user choice. Habib et al.~\cite{habib2019empirical} conducted a user study on the effectiveness of cookie consent interfaces, discovering that current implementations can violate GDPR principles by employing ``dark patterns'' to subtly nudge users towards accepting all cookies.

\textit{Readability and dark patterns.} Numerous other usability issues have been identified, such as the poor readability of privacy policies, as highlighted by Ali et al.~\cite{ali2020readability} poor readability directly affects user comprehension.
The pervasive presence of dark patterns that manipulate privacy choices has been a significant concern. Rinehart et al.~\cite{rinehart2022bringing} and Toth et al.~\cite{toth2022dark} specifically address these manipulative design practices that hinder effective user privacy control. Furthermore, opaque interfaces for cookie opt-out also contribute to these issues.

\textit{Impact of regulations and CMP adoption.} From a compliance perspective, Degeling et al.~\cite{degeling2018we} measured the contribution of implementing GDPR on improving the adoption of privacy controls, such as privacy policies and consent forms. Their findings suggested that the enforcement of GDPR coincided with a notable increase in transparency across the web. Hils et al.~\cite{hils2020measuring} analysed 161 million web pages and found that implementing GDPR and CCPA quadrupled the adoption of Consent Management Platforms (CMPs) from 2018 to 2020.  Matte et al.~\cite{matte2020cookie} 
found that 54\% have at least one suspected violation, such as pre-selected choices and no option to refuse consent.

Despite these valuable insights, a significant limitation of prior literature is its narrow scope, as it primarily focuses on one specific privacy control and fails to capture the overall effectiveness of privacy controls in terms of user experience. This limited focus can lead to situations where a website performs well in one aspect of privacy control but poorly in another. For example, a website might have a well-structured and comprehensible privacy policy, yet require five clicks to access its privacy settings, making the overall experience inefficient. 

In comparison, our study directly addresses these critical gaps in understanding the usability and implementation of privacy controls on websites. By taking a holistic approach, we empirically analyze four key privacy controls: privacy nudges, privacy notices, privacy policies, and privacy settings. Our investigation spans 18,628 websites across diverse categories and three distinct user interaction scenarios: a guest visit, a sign-up process, and an authenticated session. We apply a custom evaluation framework to assess the usability of these controls, considering five key usability principles: Awareness, Efficiency, Comprehension, Functionality, and Choice. This comprehensive approach offers a fresh perspective on privacy usability and provides valuable insights into the full spectrum of user experience with privacy controls on websites.

\section{Data Collection Methodology}
\label{sec:dcmethod}
Our data collection approach involved two main phases. In the first phase, we manually gathered data from 100 websites to inform the design of our data collection templates. These templates facilitated the creation of automated crawlers capable of accurately detecting and extracting relevant privacy control information at scale. In the second phase, we used automated web crawling scripts to systematically collect and store data from 18,628 websites while applying the same templates to ensure consistency across the dataset.

\subsection{Manual Data Collection}

\begin{table*}[htb]
\tiny
\caption{Usability attributes for privacy controls in three visit scenarios.}
\vspace{-0.4cm}
\label{table:usability}
\centering
\tabcolsep=0.05cm
\resizebox{0.8\textwidth}{!}{\begin{tabular}{llcccc|cccc|c}
\toprule
\multicolumn{2}{c}{\multirow{2}{*}{\textbf{Usability Attributes}}}	& \multicolumn{4}{c|}{\textbf{Guest Visit}}	& \multicolumn{4}{c|}{		\textbf{Sign-up User Visit}}		&	\textbf{Registered User Visit}\\
\cline{3-11}
& &\textbf{Nudges}	& \textbf{Notices}	&		\textbf{Policies}		&	\textbf{Settings} &\textbf{Nudges}	&			\textbf{Notices}	&		\textbf{Policies}		&	\textbf{Settings} &	\textbf{Account Settings}\\
\midrule\hline
\multirow{2}{*}{Awareness} 	&	Location	&	\checkmark	&	\checkmark	&	\checkmark	&	\checkmark	&	\checkmark	&	\checkmark	&	\checkmark	&	\checkmark	&	\checkmark	\\
		&	Display type	&	\checkmark	&	\checkmark	&		&		&	\checkmark	&	\checkmark	&		&	\checkmark	&		\\
	Efficiency	&	Number of clicks	&		&		&	\checkmark	&	\checkmark	&		&		&		&		&	\checkmark	\\
\multirow{2}{*}{Comprehension}	&	Language	&		&		&	\checkmark	&	\checkmark	&		&		&		&	\checkmark	&	\checkmark	\\
		&	Table of content	&		&		&	\checkmark	&		&		&		&		&		&		\\
	Functionality	&	Type of privacy content	&	\checkmark	&	\checkmark	&		&	\checkmark	&	\checkmark	&	\checkmark	&		&	\checkmark	&	\checkmark	\\
\multirow{2}{*}{Choice}&	Types of options	&	\checkmark	&		&	\checkmark	&		&	\checkmark	&		&		&		&		\\
		&	Explicit choice guidance	&	\checkmark	&		&	\checkmark	&		&	\checkmark	&		&		&		&		\\
\hline\bottomrule
\end{tabular}}
\vspace{-0.4cm}
\end{table*}

During the manual data collection phase, we performed an expert evaluation of the usability of four privacy controls across 100 health websites, selected based on their popularity and web traffic rankings from Similarweb~\cite{similarweb}. This evaluation was guided by established usability heuristics~\cite{albesher2021evaluating, redmiles2020comprehensive} and focused on five key usability attributes: \textit{awareness, efficiency, comprehension, functionality,} and \textit{choice}. This initial expert evaluation on a focused set of health websites allowed us to refine our understanding of diverse privacy control implementations and develop robust, generalizable patterns and metrics for our automated crawling scripts, rather than deriving domain-specific usability conclusions from this phase. %
Table~\ref{table:usability} provides a detailed breakdown of the usability criteria applied to each privacy control.

\begin{itemize}[left=0pt]
\item\textbf{Awareness.} refers to the degree to which users can efficiently grasp the information presented on privacy controls~\cite{albesher2021evaluating}~\cite{habib2022okay}. It assesses usability by considering factors such as visibility (\textit{display type}) and the simplicity of finding and accessing (\textit{location/placement}) the displayed information.

\item\textbf{Efficiency.} measures the time and effort users need to access privacy controls~\cite{redmiles2020comprehensive}. This attribute comprises the \textit{number of clicks} needed to access the privacy control. The ease of accessing privacy information and settings is directly related to the number of clicks.

\item\textbf{Comprehension.} refers to users' understanding and interpretation of provided information, particularly for privacy policies and settings~\cite{redmiles2020comprehensive}. It involves \textit{language} and the \textit{table of contents} as two main factors. Language relates to the clarity and readability of text in privacy controls. It emphasizes the need for easily understandable content by providing translation in multiple languages. The table of contents refers to an organized outline with hyperlinks in privacy policies or settings, which aids users in navigating lengthy texts efficiently. Note that our automated system identifies the presence of a table of contents as an indicator of structural aid for comprehension, though it does not evaluate its effectiveness or clarity in guiding users.

\item\textbf{Functionality.} evaluates the inclusiveness of privacy content within controls~\cite{redmiles2020comprehensive}. The \textit{type of privacy content} assesses the range of privacy data covered by controls, encompassing aspects like advertisements, cookies, email notifications, and other relevant topics. A comprehensive coverage of these privacy content types ensures effective communication of website data collection practices to users.

\item\textbf{Choice.} considers the influence of dark patterns on usability, focusing on meaningful user options~\cite{rinehart2022bringing}. Choice considers two vital aspects: the \textit{types of options} available to users and the provision of \textit{explicit choice guidance}.  In privacy controls like nudges, users should find meaningful options for managing their privacy preferences. Clear guidance involves concise steps to help users understand the implications of their choices.

\end{itemize}

\subsubsection{Usability evaluation in three Visit Scenarios} To assess privacy controls across different user contexts, we developed three visit scenarios—guest visits, sign-up user visits, and registered user visits. These scenarios reflect typical stages in a user's interaction with a website, allowing us to capture privacy controls as they appear during various levels of engagement. To maintain a logical flow that mirrors a natural browsing experience, the scenarios are ordered sequentially: starting with guest visits, followed by sign-up user visits (account creation), and concluding with registered user visits.
\begin{itemize}[left=0pt]
\item \textbf{Guest Visit Scenario.} It involves users interacting with a website without creating an account or logging in. Users have limited permissions and restricted access to certain content.

\item \textbf{Sign-up User Visit Scenario.} This scenario involves users creating a new account or sign up for membership on a website. The process includes providing information such as username, password, email address, and potentially additional details.

\item \textbf{Registered User Visit Scenario.} After successfully creating an account, the registered user visit scenario involves logging into the website using account credentials. In this scenario, users gain access to account information, manage settings, and experience personalized content or activities. The web crawl for the registered user visit scenario focuses on privacy nudges, privacy notices, and privacy settings. 

\end{itemize}

\subsubsection{Data Collection by Researchers} To ensure impartiality in data collection, three researchers, a post-doc, a Ph.D. student, and an undergraduate, independently completed the data collection templates. Discrepancies were reviewed collaboratively, considering factors such as browser variations, \textit{the presence of dark patterns}, and updates to privacy policies. For example, a privacy notice might appear in Chrome but not in Edge. Researchers used different browsers in incognito mode with cleared cookies to reflect realistic user experiences and avoid bias from prior visits. For account creation, mock personal data and email addresses were used, although registration sometimes failed due to missing details like a local phone number or healthcare card. When evaluating privacy policies, researchers read the full documents, and for complex websites with three user scenarios, this process could take up to 50 minutes per site. All collected data were consolidated through consensus to ensure consistency and validity in assessing the usability of privacy controls.

\subsection{Automated Data Collection via Web Crawler}

After finalizing the manual data collection template through thorough verification, we developed and deployed a web crawler to examine the privacy controls of 18,628 websites sourced from the Hispar~\cite{hispar} and SimilarWeb lists {~\cite{aqeel2020landing}}. We used Selenium-based scripts for dynamic webpage interaction and integrated the Googletrans translation API to enable language-specific analysis, allowing effective data extraction from both English and non-English websites. 

The crawler simulated user behavior across three interaction scenarios: (i) guest visits, (ii) sign-up visits, and (iii) authenticated (logged-in) sessions, automating account registration when feasible. Detailed procedures for each scenario are presented in Tables~\ref{tab:webcrawltemplateguest},\ref{tab:webcrawltemplatesignup}, and\ref{tab:webcrawltemplateregistered} in the Appendix.

Despite rigorous engineering efforts, large-scale automation of account creation and session management presented considerable challenges. For example, of the 4,070 websites identified as offering a sign-up interface, automated account registration succeeded on only 828 websites, yielding a 20.3\% success rate. This limitation primarily stemmed from technical barriers such as the requirement for credit card information, CAPTCHA complexity, and multi-factor authentication (MFA). Even during manual testing, researchers encountered cases where registration was blocked due to requirements for localized information (e.g., phone numbers or government-issued IDs). These challenges directly impacted our ability to collect sufficient data for the authenticated user scenario, resulting in a smaller and less representative dataset that constrains the generalizability of findings for logged-in user interactions.

The data collection process began by detecting the presence of privacy nudges, upon which the crawler extracted their screen location, trigger mechanism, and the type of privacy-related functionality offered (e.g., cookie preferences or data collection consent). A similar strategy was applied to other privacy controls, including privacy notices, privacy policies, and privacy settings, where presence was confirmed and relevant structural and textual details were logged.

Table~\ref{availability_scenario} summarizes the statistical availability of each privacy control across the three user visit scenarios. Notably, our analysis revealed a complete absence of privacy nudges during the registered user scenario, suggesting that websites primarily deploy such nudges during early interaction stages--namely, guest browsing or account registration--and rarely target users after authentication.

\begin{table}[htbp]
\centering
  \caption{Availability of privacy controls in three scenarios.}
  \label{availability_scenario}
  \vspace{-0.4cm}
  \small
\tabcolsep=0.05cm
\resizebox{0.46\textwidth}{!}{\begin{tabular}{l|lll}
\toprule
\multirow{2}{*}{\textbf{Privacy Control}} & \multicolumn{3}{c}{\textbf{Visit Scenarios}}                                                                                                                                                                                                                \\ \cline{2-4} 
                                          & \multicolumn{1}{l|}{\textbf{\begin{tabular}[c]{@{}l@{}}Guest (Out of)\end{tabular}}} & \multicolumn{1}{l|}{\textbf{\begin{tabular}[c]{@{}l@{}}Sign-Up (Out of)\end{tabular}}} & \textbf{\begin{tabular}[c]{@{}l@{}}Registered (Out of)\end{tabular}} \\ \hline\hline
\textbf{Privacy Nudges}                   & \multicolumn{1}{c|}{5.8\%}                                                              & \multicolumn{1}{c|}{8.6\%}                                                               &  \multicolumn{1}{c}{0\%}                                                                    \\
\textbf{Privacy Notices}                  & \multicolumn{1}{c|}{2.6\%}                                                              & \multicolumn{1}{c|}{4.0\%}                                                               &  \multicolumn{1}{c}{4.8\%}                                                                  \\
\textbf{Privacy Policies}                 & \multicolumn{1}{c|}{53.5\%}                                                             & \multicolumn{1}{c|}{76.2\%}                                                              &  \multicolumn{1}{c}{76.2\%}                                                                 \\
\textbf{Privacy Settings}                 & \multicolumn{1}{c|}{7.3\%}                                                              & \multicolumn{1}{c|}{7.2\%}                                                              &  \multicolumn{1}{c}{3.6\%}   \\           \hline \bottomrule
\end{tabular}}
\vspace{-0.3cm}
\end{table}

\subsubsection{Website Selection and Classification} To build a diverse and representative dataset, we collected 16,628 websites from the Hispar project and an archived Alexa list. To further enhance the dataset and strengthen the credibility of website rankings, we added a random sample of 2,000 websites from the Majestic Million list, specifically targeting ranks 95,000–100,000 and 900,000–1,000,000. This broader sampling enriched our analysis by covering a wider spectrum of web traffic. To validate the reliability of the rankings, we cross-referenced our dataset with Quantcast rankings. Table~\ref{tab:rankings} presents the distribution of Quantcast rankings for the scraped websites.

\begin{table}[htbp]
  \centering
  \caption{Hispar/Alexa list mapped to Quantcast rankings.}
  \vspace{-0.4cm}
  \small
  \label{tab:rankings}
  \tabcolsep=0.05cm
\resizebox{0.43\textwidth}{!}{\begin{tabular}{llll}
    \toprule
    \textbf{Quantcast Rank} & \textbf{\% of Websites}  & \textbf{Quantcast Rank} & \textbf{\% of Websites} \\
    \midrule \hline
    1-5,000             & 12.37\% &  400,001-500,000     & 94.55\% \\
    5,000-10,000        & 27.66\% &  500,001-600,000     & 95.78\% \\
    10,000-50,000       & 66.80\% &  600,001-700,000     & 97.29\% \\
    50,000-100,000      & 76.32\% &  700,001-800,000     & 98.56\% \\
    100,001-200,000     & 85.13\% &  800,001-900,000     & 99.50\% \\
    200,001-300,000     & 89.84\% &  900,001-1,000,000   & 100.00\% \\
    300,001-400,000     & 92.72\% & & \\
  
    \hline\bottomrule
  \end{tabular}}
\end{table}

We used \texttt{safedns.com} to assign a fine-grained category to each website. However, many of these categories overlap in scope. For instance, \textit{News and Media}, \textit{News and Social Media}, and \textit{Newsgroups and Message Boards} are treated as separate categories by SafeDNS, despite their thematic similarities. To streamline analysis and address category overlap, we implemented a mapping strategy that grouped 38 unique fine-grained categories into 11 broader, high-level classifications such as \textit{Technology, Business and Finance,} and \textit{Entertainment}. The high-level categories were defined based on thematic similarity, functional purpose, and user engagement patterns to ensure interpretability and analytical coherence across diverse website types. Table~\ref{tab:category-mapping} in the Appendix provides a detailed overview of this mapping. This consolidated categorization enhanced the clarity of our analysis by reducing fragmentation across similar topics, enabling more meaningful comparisons and insights across major content domains.
\vspace{-0.18cm}

\begin{table}[]
\caption{Categorical distribution of crawled websites.}
\label{category_overview}
\small
\vspace{-0.4cm}
\begin{tabular}{l|l|l|l}
\toprule
\textbf{Website Categories}   & \textbf{Count} & \textbf{Website Categories}  & \textbf{Count} \\ \hline \hline
\textbf{Technology}           & 4,489          & \textbf{Education}           & 1,274          \\
\textbf{Business \& Finance}  & 4,432          & \textbf{Lifestyle \& Sports} & 992            \\
\textbf{Entertainment}        & 2,294          & \textbf{Government}          & 742            \\
\textbf{News \& Social Media} & 2,138          & \textbf{Other}               & 653            \\
\textbf{Adult Content}        & 1,907          & \textbf{Malicious Content}   & 228            \\
\textbf{Shopping}             & 1,432          &                              &    \\           
\hline \bottomrule
\end{tabular}
\vspace{-0.4cm}
\end{table}

\section{Results}
\label{sec:results}

This section answers our key research questions (\S\ref{sec:intro}) and presents our findings on the usability of privacy controls on websites, based on the attributes defined in Table~\ref{table:usability}. Our analysis focuses on how the accessibility and implementation of these controls vary according to website ranking and categorization. Additionally, we assess the usability of privacy controls across the three user visit scenarios to capture differences in user experience at various levels of website engagement.

\subsection{Website Ranking and Privacy Control Usability}

\begin{figure}[h]
\begin{centering}
\includegraphics[width=0.7\linewidth]{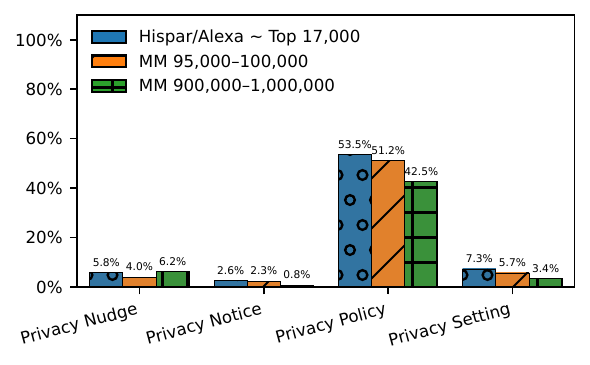}
\vspace{-0.4cm}
  \caption{Privacy control availability by website ranking.}
  \label{rankingavilability}
 \vspace{-0.4cm}
\end{centering}
\end{figure}

To answer RQ1 (\S\ref{sec:intro}), we begin by examining the availability of privacy controls across different website ranking tiers. As shown in Figure~\ref{rankingavilability}, there is a clear trend where lower-ranked websites (MM 900,000–1,000,000) tend to lack privacy controls. However, this pattern does not apply to privacy nudges. Lower-ranked websites show a higher prevalence of nudges, with 6.2\% of low-ranked sites featuring them, compared to 4.0\% for mid-ranked sites (MM 95,000–100,000) and 5.8\% for top-ranked sites (Hispar/Alexa list). Overall, the analysis indicates that the availability of privacy controls generally declines with ranking, though privacy nudges present a notable exception. This interesting deviation could suggest that lower-ranked websites, perhaps relying more on out-of-the-box CMS solutions or third-party plugins, might adopt privacy nudge implementations (such as generic cookie banners) at a higher rate, or they might be less optimized in their display strategies, leading to higher detection of these basic nudges by our crawler. This warrants further investigation into the specific technologies and motivations behind privacy control implementation at different ranking tiers.

\paragraph{Privacy Nudges and Notices}

\begin{figure*}[!t]
    \centering
    \subfigure[Location distribution.]{%
        \includegraphics[width=0.40\linewidth,height=1.5in]{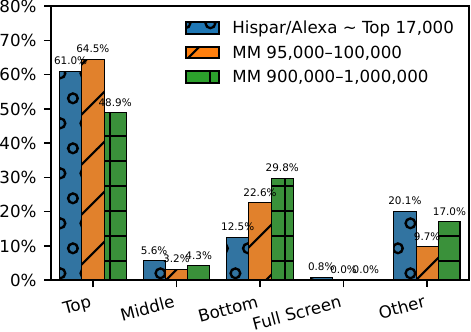} %
        \label{RankNudgeLocation}
    }\hfill
    \subfigure[Display type distribution.]{%
        \includegraphics[width=0.40\linewidth,height=1.5in]{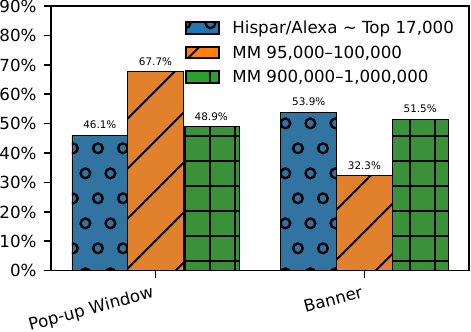}%
        \label{RankNudgeDisplayType}
    }
     \vspace{-0.4cm}
    \caption{Privacy nudges: awareness by website ranking.}
    \label{fig:privacy_nudges}
    \vspace{-0.3cm}
\end{figure*}

Figs.~\ref{RankNudgeLocation} and~\ref{RankNudgeDisplayType} highlight differences in the awareness aspect of privacy nudges across high, mid, and low-ranked websites. In terms of nudge placement (Figure~\ref{RankNudgeLocation}), the `Top' of the page is the most common location across all ranking tiers, accounting for 61.0\% of high-ranked, 64.5\% of mid-ranked, and 48.9\% of low-ranked websites. In contrast, the `Full Screen' option appears rarely--just 0.8\% in high-ranked sites and not at all in mid or low-ranked ones. This trend suggests a widespread preference for top-of-page placement to maximize user visibility and engagement, regardless of rank.

Figure~\ref{RankNudgeDisplayType} reveals a distinct pattern in how nudges are displayed, particularly in mid-ranked websites, where 67.7\% use pop-up windows—considerably more than high-ranked (46.1\%) and low-ranked (48.9\%) sites.%
The higher prevalence of pop-up windows for nudges on mid-ranked websites (67.7\%) compared to high-ranked (46.1\%) and low-ranked (48.9\%) sites is an interesting observation. Future qualitative research could explore whether this display strategy is employed to enhance user engagement or awareness on sites with potentially lower inherent user trust or brand recognition.

\begin{figure}[h]
\begin{centering}
\includegraphics[width=0.43\textwidth, height=1.5in]{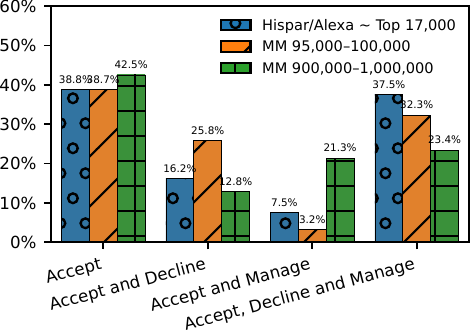}
 \vspace{-0.4cm}
  \caption{Privacy nudge options by website ranking.}
  \label{RankNudgeOpTypes}
\end{centering}
\vspace{-0.3cm}
\end{figure}

An analysis of privacy nudge options in Figure~\ref{RankNudgeOpTypes} shows that top and mid-ranked websites exhibit similar distributions, offering a balanced range of choices: `accept', `accept and decline', and `accept, decline, and manage'. In contrast, low-ranked websites offer fewer options, with a notably higher proportion (42.5\%) restricting users to only accepting the nudge.

\begin{figure}[h]
\begin{centering}
\includegraphics[width=0.43\textwidth, height=1.5in]{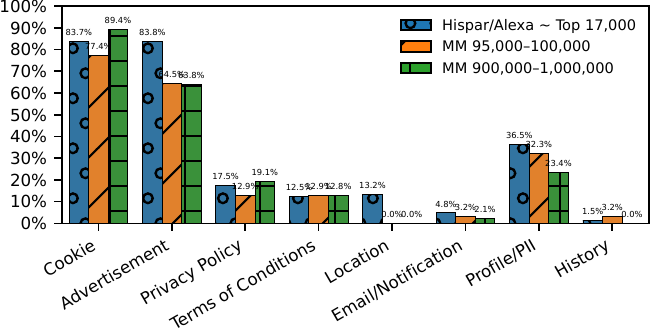}
 \vspace{-0.4cm}
  \caption{Privacy nudge content types by website ranking.}
  \label{RankNudgeContentType}
\end{centering}
\vspace{-0.4cm}
\end{figure}

Figure~\ref{RankNudgeContentType} provides a detailed distribution of nudge functionality across the three website ranking groups. Notably, nudges containing cookie and privacy policy information are most common on low-ranked websites (83.7\%), followed by high-ranked sites (89.4\%), with mid-ranked sites showing the lowest frequency (77.4\%). For content types related to advertisements, email/notifications, and profile/PII, a clear descending trend is seen, with significantly higher occurrences on high-ranked websites. Terms and conditions and history nudges, however, are distributed fairly evenly across all ranking categories. Interestingly, mid- and low-ranked websites do not feature nudges related to location information, which could be due to the smaller sample size of 1,000 websites in these categories.

\begin{figure}[h]
\begin{centering}
\includegraphics[width=0.4\textwidth, height=1.5in]{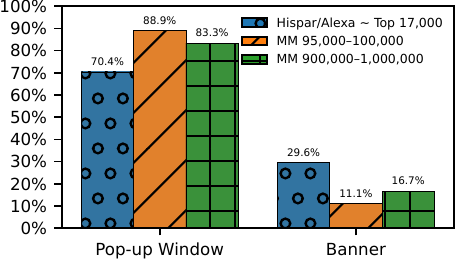}
  \caption{Notice display type distribution by website ranking.}
  \label{RankNoticeDisplayType}
\end{centering}
\vspace{-0.4cm}
\end{figure}

Compared to privacy nudges, privacy notices are predominantly displayed as `Pop-up Windows'. While pop-ups can be seen as intrusive and disruptive to the browsing experience, when designed effectively, they can serve as a powerful tool to inform users about their privacy rights and enhance privacy awareness. As shown in Figure~\ref{RankNoticeDisplayType}, both mid- and low-ranked websites follow a similar trend, with around 85\% of notices displayed as pop-ups. In contrast, high-ranked websites have a slightly lower percentage, at 70.4\%. This difference suggests that lower and mid-ranked websites may prioritize pop-up notices to grab user attention, while high-ranked websites may reduce their use of pop-ups to avoid disrupting the user experience.

\paragraph{Privacy Policies and Settings}

\begin{figure*}[!t]
    \centering
    \subfigure[Location distribution for policy.]{%
        \includegraphics[width=0.43\linewidth, height=1.5in]{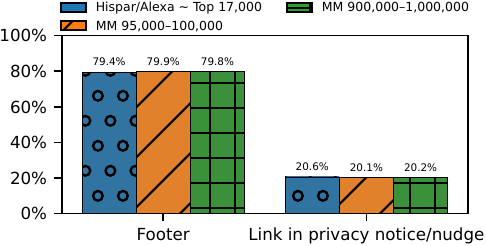} %
        \label{RankPolicyLocation}
    }\hfill
    \subfigure[Clicks to access policy.]{%
        \includegraphics[width=0.43\linewidth, height=1.5in]{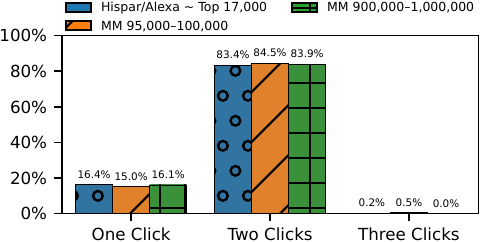}%
        \label{RankPolicyClicks}
    }
     \vspace{-0.4cm}
    \caption{Privacy policy: awareness and efficiency by ranking.}
    \label{fig:privacy_policies}
    \vspace{-0.3cm}
\end{figure*}

In analyzing the awareness attribute of privacy policies, we focus on their location on websites. Figure~\ref{RankPolicyLocation} shows a consistent pattern across high-, mid, and low-ranked websites, with roughly 80\% placing privacy policies in the `Footer'. A smaller share, about 20\%, appears as a `Link in privacy notice/nudge'. This indicates a clear preference across all ranking tiers for positioning privacy policies in the footer, reflecting a widely adopted convention to ensure users can reliably access privacy information.

To assess the efficiency of accessing privacy policies, we measured the number of clicks required. Most websites require two clicks to reach the privacy policy via a button or link. As shown in Figure~\ref{RankPolicyClicks}, top-ranked websites offer slightly better access, with 16.4\% requiring only one click, compared to 15.0\% for mid-ranked and 16.1\% for low-ranked sites. This indicates a small but notable edge in user accessibility on higher-ranked websites, with room for improvement among lower-ranked ones.

\begin{figure}[h]
\begin{centering}
\includegraphics[width=0.43\textwidth, height=1.5in]{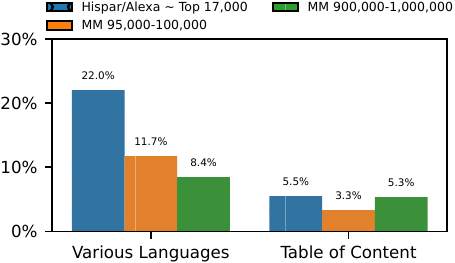}
 \vspace{-0.4cm}
  \caption{Policies with languages and ToC across ranking.}
  \label{RankPolicyComprehension}
\end{centering}
\vspace{-0.3cm}
\end{figure}

For comprehension, we examined whether the privacy policy is available in multiple languages and whether it includes a table of contents. Figure~\ref{RankPolicyComprehension} shows a clear decline in multilingual support with decreasing rank; 22\% of top-ranked sites offer multiple languages, compared to just 8.4\% of low-ranked ones. This suggests top-ranked websites are more attentive to serving diverse user bases. In contrast, the inclusion of a table of contents shows minor variation: it drops from 5.5\% (top-ranked) to 3.3\% (mid-ranked), then slightly increases to 5.3\% for low-ranked sites. Overall, higher-ranked websites emphasize accessibility through language support, while structural aids like tables of contents are less consistently implemented across rankings. 

\begin{figure}[h]
\begin{centering}
\includegraphics[width=0.43\textwidth, height=1.5in]{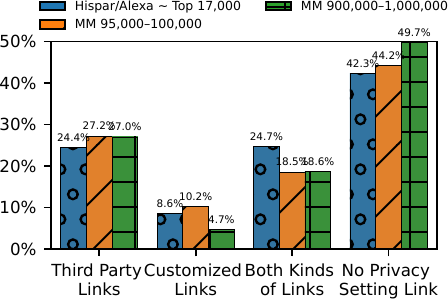}
 \vspace{-0.4cm}
  \caption{Privacy setting link types by ranking.}
  \label{RankPolicyLinkType}
\end{centering}
\vspace{-0.3cm}
\end{figure}

Figure~\ref{RankPolicyLinkType} shows a clear decline in the availability of privacy setting links within privacy policies as website ranking decreases. While 57.7\% of top-ranked sites include such links, the percentage drops to 55.8\% for mid-ranked and 50.3\% for low-ranked websites. This trend suggests that higher-ranked websites are more likely to provide users with direct access to privacy settings. Moreover, top-ranked websites also lead in the quality of these links, with 24.7\% offering both customized and third-party options. In comparison, mid- and low-ranked sites provide this combined option at lower rates—around 18\%. This indicates that higher-ranked websites not only offer more frequent access to privacy settings but also deliver more comprehensive choices, reflecting stronger support for the `Choice' usability attribute.

\begin{figure}
\centering  
\subfigure[Location Distribution for Setting]{\label{RankSettingLocation}\includegraphics[width=0.236\textwidth, keepaspectratio]{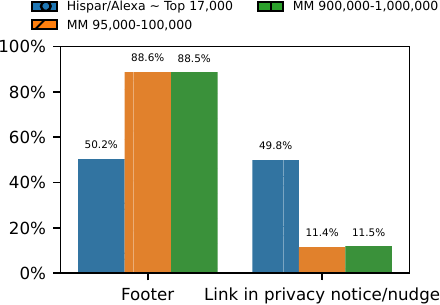}}
\subfigure[Clicks to access Setting]{\label{RankSettingClicks}\includegraphics[width=0.236\textwidth, keepaspectratio]{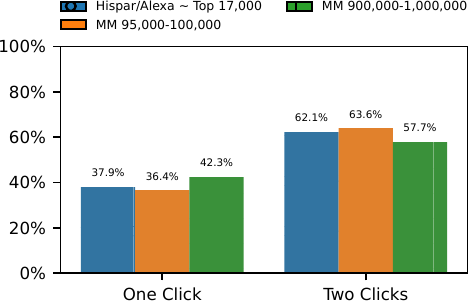}}
 \vspace{-0.4cm}
\caption{Privacy setting: awareness and access efficiency.}
\vspace{-0.4cm}
\end{figure}

As illustrated in Figure~\ref{RankSettingLocation}, middle and low-ranked websites primarily place privacy setting links in the footer. In contrast, top-ranked websites adopt a more balanced approach—50.2\% place the link in the footer, while 49.8\% embed it within the notice or nudge, indicating an effort to enhance visibility. Regarding access efficiency, top and mid-ranked websites show comparable performance, with about 37.9\% allowing users to reach settings in a single click. Interestingly, low-ranked websites perform slightly better, with 42.3\% offering one-click access (Figure~\ref{RankSettingClicks}). 

The finding that low-ranked websites show slightly better one-click access to privacy settings (42.3\%) compared to top/mid-ranked sites (37.9\%) is noteworthy. This could be attributed to simpler site architectures on lower-ranked sites, where privacy settings might be more directly exposed, or a less complex user interface overall, reducing the number of clicks required to reach specific elements. In contrast, higher-ranked sites might integrate settings within more intricate navigation menus, inadvertently increasing click-depth.

\subsection{Website Category and Privacy Control Usability}

By exploring the influence of website categories on the usability of privacy controls, in this section, we answer our  RQ2 (\S\ref{sec:intro}). 

It is critical to highlight that due to the significantly small sample size of successfully collected data for the registered visit scenario, comprehensive categorical insights for this particular scenario are not provided. Therefore, this analysis primarily focuses on standard website visits (guest user visits), with additional insights drawn from the sign-up visit scenario where sufficient data allowed.

\subsubsection{Privacy Nudges} 
\begin{figure*}
\centering  
\subfigure[Location distribution.]{\label{CatNudgeLocation}\includegraphics[width=0.43\textwidth, height=1.5in]{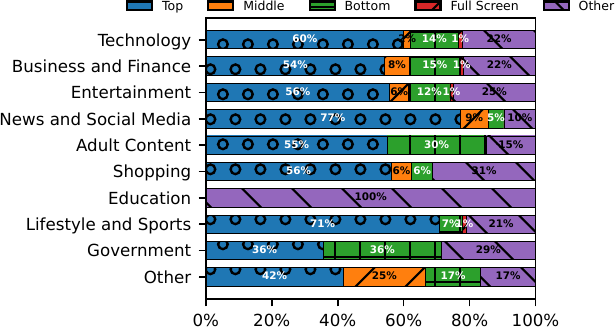}
}\hfill
\subfigure[Display type distribution.]{\label{CatNudgeDisplayType}\includegraphics[width=0.43\textwidth, height=1.5in]{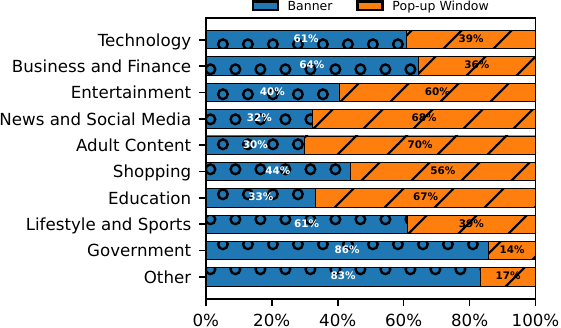}}
 \vspace{-0.4cm}
\caption{{Privacy nudge awareness across website categories.}}
\vspace{-0.2cm}
\end{figure*}

Figure~\ref{CatNudgeLocation} shows that the majority of website categories prefer placing privacy nudges at the top of the page. However, some categories show variation. For instance, 35.7\% of nudges on government websites are placed at the bottom, while Education websites exclusively position them elsewhere. Overall, the widespread placement of nudges at the top of the page across categories suggests a strategic emphasis on capturing user attention, with the bottom placement being the next most common choice.

Figure~\ref{CatNudgeDisplayType} provides a detailed breakdown of nudge display types across website categories. Banners are the most common display type in categories such as Technology (60.9\%), Business and Finance (64.4\%), Lifestyle and Sports (61.1\%), Government (85.7\%), and Other (83.3\%). Shopping websites show a more balanced distribution, with banners slightly ahead at 56.25\%. In contrast, pop-up windows are more prevalent in categories like Entertainment (59.6\%), News and Social Media (67.6\%), Education (66.7\%), and Adult Content (70\%). The distribution of nudge display types varies across categories, with many using a combination of banners and pop-ups to achieve specific goals, whether it's to minimize intrusiveness or to increase user awareness through more noticeable displays.

\begin{figure}[h]
\begin{centering}
\includegraphics[width=0.43\textwidth, height=1.5in]{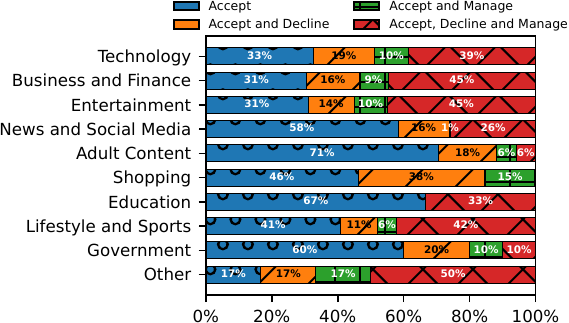}
 \vspace{-0.4cm}
  \caption{{Privacy nudge option types across web categories.}}
  \label{CatNudgeOpType}
\end{centering}
\vspace{-0.2cm}
\end{figure}

We also examined the distribution of privacy nudge option types across website categories, as shown in Figure~\ref{CatNudgeOpType}. The Technology, Business and Finance, and Entertainment categories exhibit similar patterns, with a fairly even spread among `accept', `accept and decline', and `accept, decline, and manage' options. In contrast, the Adult Content category stands out with a strong preference (70.6\%) for nudges that allow users only to accept. Similarly, in the Education category, about two-thirds of websites offer only the `Accept' option. While offering just the `Accept' option simplifies the decision-making process, it may restrict user autonomy, as it forces users to accept terms without alternatives. On the other hand, the `Manage' option allows for more personalized choices, giving users greater flexibility. Overall, the distribution of privacy nudge option types reveals a strategic approach to balancing user empowerment with simplified decision-making.

The analysis of privacy nudges by website category highlights distinct patterns in the distribution of privacy content types. The eight examined functionalities, \textit{cookie, advertisement, privacy policy, terms and conditions, location, email/notification, personally identifiable information (PII),} and \textit{history}, vary in emphasis across categories. Cookies and privacy policies are the most frequently addressed. As shown in Table~\ref{tab:functionality-analysis}, \textit{Shopping} websites prioritize `Advertisement' content, with full coverage (100\%), indicating a strong focus on ad transparency and user targeting.\textit{Government} websites stand out for their emphasis on location-related content (35.7\%), reflecting the relevance of geolocation in public services. These variations suggest that websites tailor their privacy disclosures based on the nature of their services and expected user interactions.

\begin{table*}[htbp]
\scriptsize
    \centering
    \caption{Functionality analysis of privacy nudges across website categories.}
    \label{tab:functionality-analysis}
    \vspace{-0.4cm}
    \resizebox{0.90\textwidth}{!}{%
        \begin{tabular}{l|p{1.2cm}|p{1.2cm}|p{1.2cm}|p{1.2cm}|p{1.2cm}|p{1.2cm}|p{1.2cm}|p{1.2cm}}
        \toprule
        \textbf{Category } & \textbf{Nudge Count} & \textbf{Cookie} & \textbf{Advertisement} & \textbf{Privacy Policy} & \textbf{T\&C} & \textbf{Location} & \textbf{ Email/ Notification} & \textbf{Profile/ PII} \\
        \hline\hline
        Technology & 230 & 80.9\% & 80.4\% & 12.6\% & 10.0\% & 8.7\% & 6.1\% & 35.2\% \\
        Business and Finance & 306 & 84.3\% & 79.4\% & 18.9\% & 8.5\% & 7.2\% & 5.6\% & 40.8\% \\
        Entertainment & 104 & 90.4\% & 89.4\% & 17.3\% & 10.6\% & 10.6\% & 2.9\% & 49.0\% \\
        News and Social Media & 188 & 90.4\% & 90.4\% & 17.6\% & 18.6\% & 27.7\% & 2.1\% & 26.6\% \\
        Adult Content & 20 & 35.0\% & 90.0\% & 15.0\% & 25.0\% & 0.0\% & 10.0\% & 10.0\% \\
        Shopping & 16 & 68.8\% & \textbf{100.0\%} & 18.8\% & 18.8\% & 6.3\% & \textbf{12.5\%} & 18.8\% \\
        Education & 3 & \textbf{100.0\%} & 66.7\% & 0.0\% & \textbf{33.3\%} & 0.0\% & 0.0\% & 33.3\% \\
        Lifestyle and Sports & 72 & 80.6\% & 88.9\% & 22.2\% & 22.2\% & \textbf{27.8\%} & 4.2\% & 43.1\% \\
        Government & 14 & 64.3\% & 71.4\% & \textbf{35.7\%} & 7.1\% & 7.1\% & 7.1\% & 7.1\% \\
        Other & 12 & 100.0\% & 66.7\% & 33.3\% & 0.0\% & 0.0\% & 0.0\% & \textbf{58.3\%} \\
        Malicious Content & 0 & 0.0\% & 0.0\% & 0.0\% & 0.0\% & 0.0\% & 0.0\% & 0.0\% \\
        \hline \hline
    \end{tabular}%
    }
\end{table*}

\subsubsection{Privacy Notices} 
\begin{figure}[h]
\begin{centering}
\includegraphics[width=0.43\textwidth, height=1.5in]{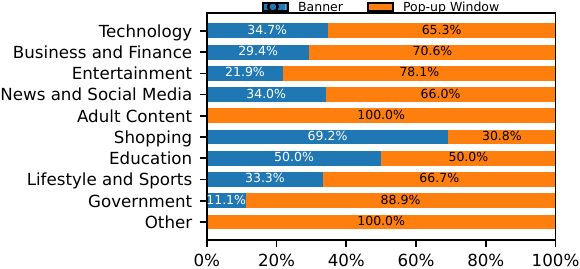}
 \vspace{-0.4cm}
  \caption{Display type of privacy notices by website category.}
  \label{CatNoticeDisplayType}
\end{centering}
\vspace{-0.4cm}
\end{figure}

Figure~\ref{CatNoticeDisplayType} illustrates how privacy notice display types vary across website categories. Most categories show a strong preference for pop-up windows. This is especially true for the \textit{Adult Content} and \textit{Other} categories, where pop-ups account for 100\% of displays. Conversely, \textit{Education and Technology} shows a more balanced approach, with banner notices making up 50\% and 34.7\% respectively. The \textit{Government} category heavily favors pop-ups (88.9\%), indicating a more forceful communication style, while the \textit{Shopping} category stands out by preferring banner notices (69.2\%), suggesting a less intrusive approach.

\subsubsection{Privacy Policies} 
\begin{figure}[h]
\centering  
\subfigure[Location distribution for policy.]{\label{CatPolicyLocation}\includegraphics[width=0.343\textwidth, height=1.25in]{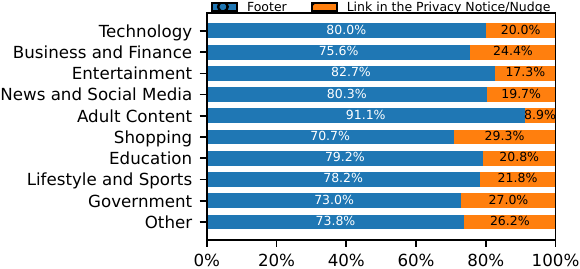}
}\hfill
\subfigure[Clicks to access policy.]{\label{CatPolicyClicks}\includegraphics[width=0.343\textwidth,height=1.25in]{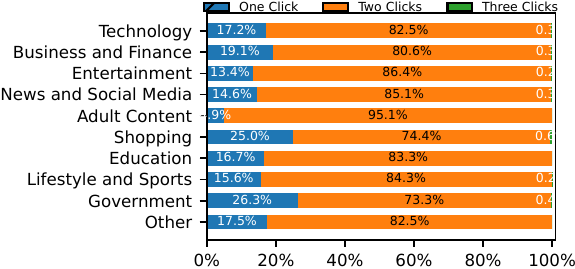}}
 \vspace{-0.4cm}
\caption{Privacy policy: awareness and efficiency across categories.}
\vspace{-0.4cm}
\end{figure}

To assess the \textit{awareness} attribute of privacy policies, we examine their placement on websites. As shown in Figure~\ref{CatPolicyLocation}, most categories position the privacy policy link in the footer. Notably, \textit{Shopping} websites lead in placing policies within a privacy nudge or notice (29.3\%), followed by \textit{Government} sites (27.0\%). For efficiency, measured by the \textit{number of clicks} needed to access the policy, Figure~\ref{CatPolicyClicks} reveals a common two-click pattern across categories. However, \textit{Government} and \textit{Shopping} sites stand out for their relatively higher proportion of one-click access (26.3\% and 25.0\%, respectively). These trends highlight consistent efforts across categories to enhance policy accessibility and transparency in data practices.

\begin{figure}[h]
\centering  
\subfigure[Policies with various languages.]{\label{CatPolicyLanguage}\includegraphics[width=0.343\textwidth,height=1.25in]{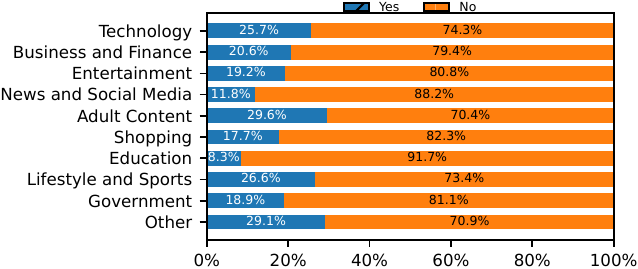}
}\hfill
\subfigure[Policies with table of contents.]{\label{CatPolicyToC}\includegraphics[width=0.343\textwidth, height=1.25in]{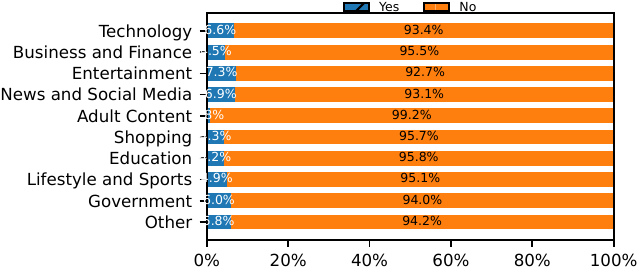}}
 \vspace{-0.4cm}
\caption{Language and ToC presence in policies by category.}
\label{CatPolicyComprehension}
\vspace{-0.4cm}
\end{figure}

To evaluate how easily users can understand privacy policies, we examined whether the policies are offered in multiple languages and include a table of contents. As shown in Figure~\ref{CatPolicyComprehension}, most website categories lack these features. Multilingual availability is generally low, varying from 8.3\% in \textit{Education} to 29.6\% in \textit{Adult Content}, indicating limited language accessibility. Similarly, fewer than 10\% of websites in any category include a table of contents, which could hinder navigation and comprehension. This absence is particularly notable in \textit{Government} and \textit{Technology} websites, where privacy policies tend to be longer and more complex--making a structured outline especially valuable for users seeking specific information.

\begin{figure}[h]
\centering  
\subfigure[Privacy setting type distribution.]{\label{CatPolicySetLinkType}\includegraphics[width=0.343\textwidth, height=1.25in]{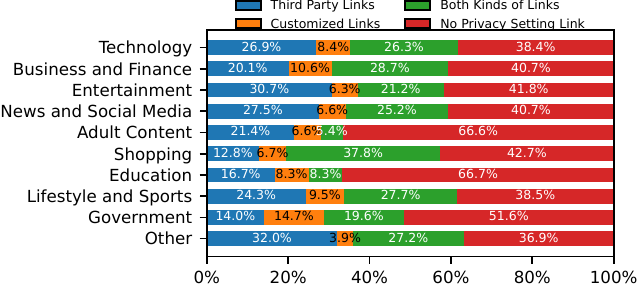}}
\subfigure[Policies with clear setting.]{\label{CatPolicyClearGuide}\includegraphics[width=0.343\textwidth, height=1.25in]{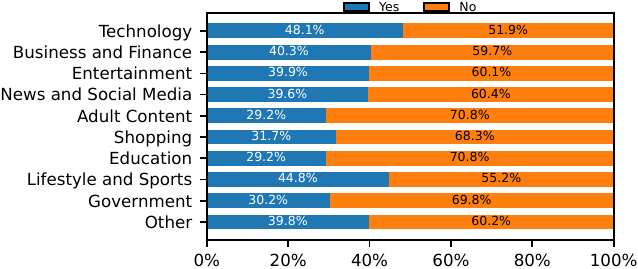}}
 \vspace{-0.4cm}
\caption{Privacy policy: choice attribute analysis across website categories.}
\vspace{-0.4cm}
\end{figure}

The type of option refers to the range of privacy settings or choices provided within a website's privacy policy, allowing users to make informed decisions about their personal data. As shown in Figure~\ref{CatPolicySetLinkType}, a significant portion of websites across categories, ranging from 38.4\% in \textit{Technology} to 66.7\% in \textit{Education}, offer no direct links to privacy settings. Additionally, about one-fifth of websites across all categories rely solely on external resources like `All About Cookies', `Your Online Choices', or third-party privacy pages (e.g., Adobe, Google) instead of offering tailored in-policy links. Notably, 37.8\% of \textit{Shopping} websites include both third-party and customised privacy setting links, highlighting better integration. These insights emphasize the need for user-friendly, embedded privacy setting options within policies to improve accessibility and usability. Even when custom links are absent, providing clear guidance, as seen in the \textit{Technology} (48.1\%) and \textit {Lifestyle \& Sports} (44.8\%) categories, per Figure~\ref{CatPolicyClearGuide}--is a positive step toward respecting user privacy choices.

\subsubsection{Privacy Setting}
\begin{figure}[h]
\centering  
\subfigure[Location distribution for setting.]{\label{CatSettingLocation}\includegraphics[width=0.343\textwidth, height=1.3in]{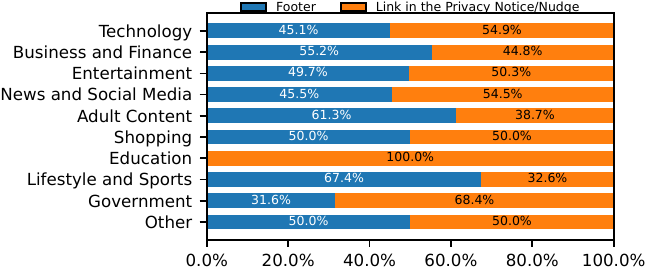}}

\subfigure[Clicks to access setting.]{\label{CatSettingClicks}\includegraphics[width=0.343\textwidth, height=1.3in]{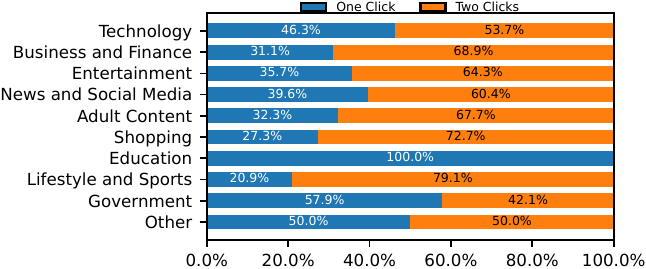}}
 \vspace{-0.4cm}
\caption{Privacy settings: awareness and efficiency across categories.}
\vspace{-0.3cm}
\end{figure}

As shown in Figure~\ref{CatSettingLocation}, the display locations of the privacy setting show a more balanced distribution between the footer and the link in privacy nudge or notice.

\subsection{Usability of Privacy Controls across three Visit Scenarios}

In this section, we answer our RQ3 (\S\ref{sec:intro}). In essence, we evaluate the usability of privacy controls across three user visit scenarios: \textit{guest, sign-up, and registered visits.}

\subsubsection{Privacy Nudges}

\begin{figure}
\centering  
\subfigure[Location distribution.]{\label{VisitNudgeLocation}\includegraphics[width=0.235\textwidth, height=1.0in]{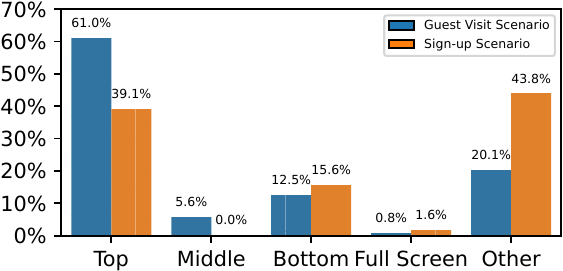}}\hfill
\subfigure[Display distribution.]{\label{VisitNudgeDisplayType}\includegraphics[width=0.235\textwidth, height=1.0in]{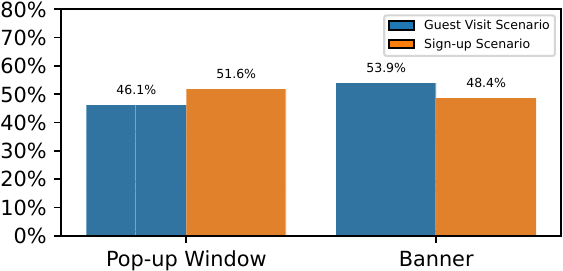}}
 \vspace{-0.4cm}
\caption{Privacy nudge awareness across visit scenarios.}
\vspace{-0.4cm}
\end{figure}

{\it Awareness.} To evaluate the effectiveness of privacy nudge presentation and placement, we examine their location and display format across different user visit scenarios. As depicted in Figure~\ref{VisitNudgeLocation}, nudges are most frequently positioned at the top of the page during guest visits (61\%), with a smaller share appearing in other areas (20.1\%). In contrast, the sign-up scenario shows a more diversified layout, with top-placement accounting for 39.1\% and `Other' positions increasing to 43.8\%. This suggests that while top placement remains a preferred strategy, sign-up processes often incorporate more flexible positioning. In terms of display type (Figure~\ref{VisitNudgeDisplayType}), guest visits exhibit a slight preference for banner-style nudges (53.9\%), whereas sign-up visits lean marginally towards pop-up windows (51.6\%). These findings highlight a balanced adoption of both display formats, reflecting an effort by Hispar/Alexa websites to align with diverse user preferences and optimize engagement across different interaction points.

\begin{figure}[h]
\begin{centering}
\includegraphics[width=0.34\textwidth, height=1.25in]{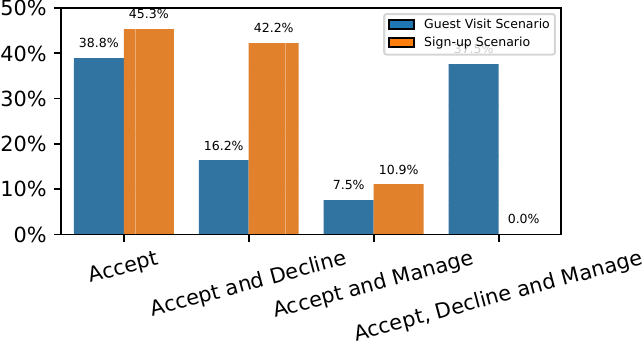}
 \vspace{-0.4cm}
  \caption{Privacy nudge option types across guest and sign-up scenarios.}
  \label{VisitNudgeOpTypes}
\end{centering}
\vspace{-0.4cm}
\end{figure}

{\it Choice.}  
In the guest visit scenario, as illustrated in Figure~\ref{VisitNudgeOpTypes}, a notable 38.8\% of privacy nudges provide only the `Accept' option, while 16.2\% include both `Accept' and 'Decline'. A smaller proportion (7.5\%) offer the combination of `Accept' and `Manage', whereas 37.5\% present users with the full set of options: `Accept', `Decline', and `Manage'. In contrast, the sign-up scenario shows a reduction in option diversity. Here, 45.3\% of nudges offer only the `Accept' option, and 42.2\% offer both `Accept' and `Decline'. A limited 10.9\% include `Accept' and `Manage', with no nudges presenting all three choices. Despite these differences, the majority of nudges in both scenarios provide explanatory information about the available options, reaching 94.5\% in the guest visit and 98.4\% in the sign-up scenario, reflecting a strong emphasis on user awareness.

In summary, a significant portion of websites overlook user choice rights in both guest and sign-up scenarios. The guest visit scenario demonstrates relatively better performance, offering users more rights compared to the sign-up scenario. Nonetheless, there is room for improvement in enhancing transparency and user control across both scenarios.

\begin{figure}[h]
\begin{centering}
\includegraphics[width=0.38\textwidth, height=1.3in]{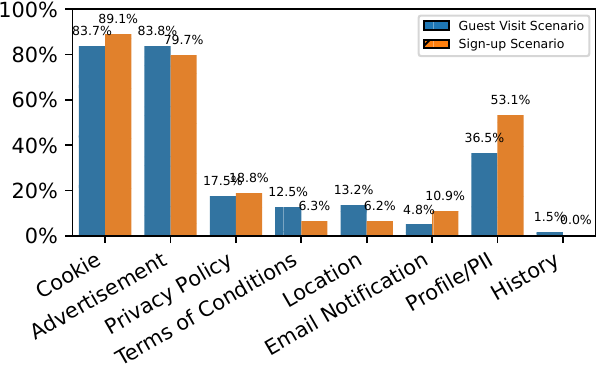}
 \vspace{-0.4cm}
  \caption{Privacy nudge: content type across guest and sign-up scenarios.}
  \label{VisitNudgeContentType}
\end{centering}
\vspace{-0.4cm}
\end{figure}

{\it Functionality.} 
Figure~\ref{VisitNudgeContentType} presents the distribution of privacy content types across guest and sign-up visit scenarios. In guest visits, privacy nudges primarily emphasize cookies (83.7\%) and advertisements (83.8\%), with privacy policies appearing in 17.5\% of cases. Other categories—such as email/notifications, PII, and browsing history—show varied coverage, with profile/PII content being the most prominent at 36.5\%. In the sign-up scenario, cookies (89.1\%) and advertisements (79.7\%) remain dominant, while privacy policy mentions rise slightly to 18.8\%. Compared to guest visits, sign-up nudges focus more on PII and do not address browsing history at all.

Overall, cookies are consistently prioritized in both scenarios. Guest visits feature a broader range of content types, whereas sign-up scenarios show a stronger emphasis on personal data (PII).

\subsubsection{Privacy Notices}

{\it Awareness.} We assess the effectiveness of privacy notices by examining their location and display type about the awareness usability attribute. Most websites do not display a privacy notice across visit scenarios. For those that do in the guest visit scenario, the majority place the notice at the bottom of the page (87.8\%) and use a pop-up window format (70.4\%). Banners are the second most common display type, appearing in 29.6\% of cases (Figure~\ref{RankNoticeDisplayType}). These patterns indicate a strong preference for specific notice placements and formats.

\begin{figure}[h]
\begin{centering}
\includegraphics[width=0.38\textwidth, height=1.3in]{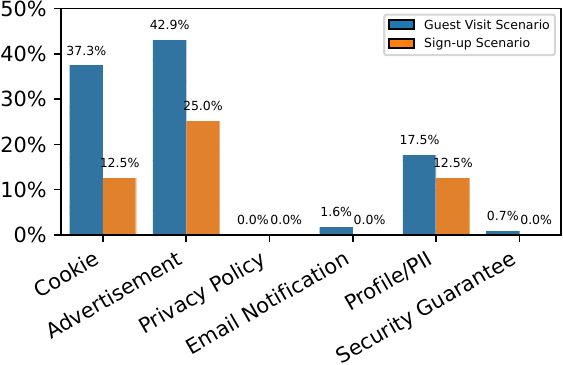}
 \vspace{-0.4cm}
  \caption{Privacy notice: content type distribution in visit scenarios.}
  \label{VisitNoticeContentTyoe}
\end{centering}
\vspace{-0.4cm}
\end{figure}

{\it Functionality.} Compared to privacy nudges, privacy notices encompass a narrower range of content types. Figure~\ref{VisitNoticeContentTyoe} illustrates this distribution. In the guest visit scenario, privacy notices primarily include content on Cookies (37.3\%), Advertisements (42.9\%), Profile/PII (17.5\%), and Email/Notifications (1.6\%). Notably, Privacy Policy content is absent, while a Security Guarantee appears uniquely at 0.7\%. In the sign-up scenario, privacy notices contain only three content types: Advertisements (25.0\%), Profile/PII (12.5\%), and Cookies (12.5\%). This contrast highlights a shift in content emphasis between guest and sign-up visits.

Our evaluation of privacy notices reveals a significant absence of such notices on the majority of websites across all visit scenarios. For those websites that do include privacy notices during guest visits, the `Bottom' display location and `Pop-up Window' display type are predominant. Additionally, privacy notice content types diverge between scenarios, with the guest visit featuring more variety, including the Security Guarantee.

\subsubsection{Privacy Policies}

\begin{figure}[h]
\begin{centering}
\includegraphics[width=0.38\textwidth, height=1.3in]{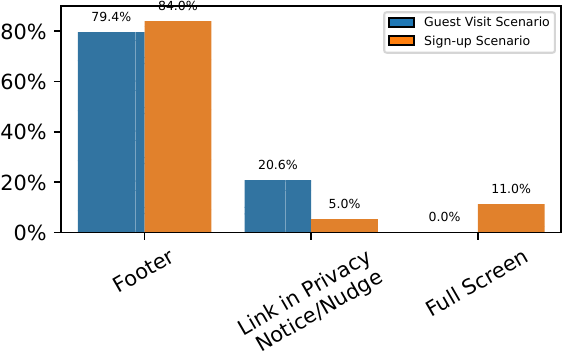}
 \vspace{-0.4cm}
  \caption{Privacy policy: location distribution in visit scenarios.}
  \label{VisitPolicyLocation}
\end{centering}
 \vspace{-0.4cm}
\end{figure}

{\it Awareness.}
As shown in Figure ~\ref{VisitPolicyLocation}, the distribution of privacy policy link locations indicates that footer links are predominant in both the guest (79.4\%) and sign-up (84.0\%) visit scenarios. Interestingly, full-screen privacy policies are found in the sign-up scenario (11.0\%), but are absent in the guest visit scenario. From an awareness perspective, placing the privacy policy link in the footer may reduce its visibility, as users must scroll to access it. However, this conventional placement meets user expectations, and while it might lower immediate visibility, it could enhance awareness due to users' familiarity with finding privacy policies in the footer.

\textit{Efficiency.} Our analysis showed that 83.4\% of top-ranked websites allow access to the privacy policy within two clicks, indicating an overall efficient process. Only a small percentage (0.2\%) of websites require three clicks, highlighting their efforts to ensure the privacy policy is easily accessible for users who actively search for it.

\textit{Comprehension.} Our findings reveal that the majority of Hispar/Alexa websites lack robust comprehension features, with only 22.0\% offering privacy policies in various languages and a mere 5.5\% incorporating a table of contents. This deficiency often results in users having to scroll through the entire privacy policy, hindering efficient navigation. Additionally, even when a table of contents is present, the lack of language options within it requires users to manually translate the content. This may pose potential challenges for those not proficient in the displayed language.

\textit{Choice.} The `type of option' refers to the various privacy settings or choices available to users within the privacy policy. Our analysis reveals that 42.3\% of websites do not provide any privacy setting links. A smaller proportion (6.4\%) exclusively offers third-party links to informational privacy setting websites. Additionally, 8.6\% provide customized links to other pages within the same domain, and 24.7\% offer both types of links. Although many websites include links related to privacy settings, a significant percentage that no links exist (42.3\%) is particularly concerning. Furthermore, 58.5\% of websites do not provide clear guidance on how to use these settings, which may undermine their effectiveness for users.

\subsubsection{Privacy Settings}

As Awareness and Efficiency attributes have been addressed in the ranking analysis, our focus is now on evaluating the comprehension and functionality of privacy settings.

{\it Comprehension.} Similar to privacy policies, most websites (78.4\%) do not offer settings in multiple languages. This distribution closely mirrors that of privacy policies, with only 21.6\% providing this feature in the guest visit scenario. This pattern highlights the need to improve user comprehension by incorporating language diversity in both privacy policies and settings.

{\it Functionality.} Cookies are the most prominent privacy content type in settings, making up a significant 22.8\%, highlighting the importance of offering cookie management options, likely due to regulatory requirements. The remaining share consists of settings that do not fall into the specified categories. Notably, email/notification settings make up a larger portion (13.7\%) in settings compared to their representation in nudges and notices, reflecting websites' focus on providing users with more control over email preferences. This aligns with GDPR and CCPA regulations, where email opt-outs are a critical component.

\section{Discussion}
Our empirical analysis reveals pervasive usability challenges that significantly hinder users' ability to effectively manage their privacy online. The results demonstrate usability flaws encountered by users, stemming from inconsistent design, accessibility issues, and a general lack of user-centric implementation of privacy controls. These findings move beyond mere technical observations to highlight critical impediments to user agency and informed decision-making in the digital sphere.

\textbf{Impeding user awareness and efficiency.} A fundamental hurdle for users is the sheer difficulty in discovering and accessing privacy controls. Our findings indicate that privacy nudges were present on only 5.8\% of top-ranked websites, while notices were found on a mere 2.6\% of guest visits to high-ranked websites. This scarcity means that many users are simply not made aware of privacy-related prompts or disclosures, thereby undermining the \textit{Awareness} usability principle, which assesses the degree to which users can efficiently grasp information on privacy controls.

Furthermore, even when privacy controls are present, accessing them often requires considerable user effort, adversely affecting overall \textit{Efficiency}. For instance, 83.4\% of top-ranked websites still require at least two clicks to access their privacy policy, while 42.3\% fail to provide any direct link to privacy settings. Users are thus forced to navigate through multiple layers of menus, echoing previously reported issues of obfuscated or poorly structured links that hinder privacy decision-making~\cite{habib2019empirical}. These inefficiencies contribute to increased user frustration and lower engagement, as the effort involved often outweighs the perceived benefits--effectively burying privacy controls.

\textbf{Undermining user comprehension.} Even when privacy controls are accessible, understanding them poses a significant challenge, directly impacting \textit{Comprehension}. Only 22\% of top-ranked websites offer multilingual privacy policies, with this figure dropping to 8.4\% for low-ranked sites. Similarly, fewer than 10\% of websites in any category include a table of contents, and just 5.5\% of top sites offer this navigational aid. These shortcomings force users to scroll through lengthy documents and manually translate content, creating barriers for non-native speakers. As a result, users struggle to understand privacy terms and implications, limiting informed decision-making. Compounding this issue, 58.5\% of websites offer no clear guidance on how to use privacy settings, further reducing their effectiveness.

\textbf{Restricting user choice and control.}
One of the most critical usability challenges involves \textit{Choice}, particularly where dark patterns limit meaningful user options. Our findings show that user autonomy is often constrained. For example, in the ``Adult Content'' and ``Education'' categories, most privacy nudges (70.6\% and 66.7\%, respectively) present only an \textit{Accept} option—effectively coercing consent. Additionally, a substantial number of websites—ranging from 38.4\% in Technology to 66.7\% in Education—offer no direct links to privacy settings within their privacy policies. About one-fifth redirect users to third-party pages, further complicating access. Overall, 42.3\% of sites lack any accessible privacy settings, undermining users' ability to make informed choices. These patterns reflect a systemic disregard for user control in both guest and sign-up scenarios.

\textbf{Inconsistent functionality and varying user experiences.}
The \textit{Functionality} of privacy controls, evaluating the inclusiveness of privacy content, also presents inconsistencies. While content types like cookies and privacy policies are frequently addressed in nudges, the range of information covered varies significantly across website categories and visit scenarios. For example, Shopping websites prioritize `Ads' content, with full coverage (100\%), indicating a strong focus on ad transparency and user targeting, while Government websites stand out for their emphasis on location-related content (35.7\%). Such variations, while sometimes tailored, contribute to a lack of standardization that can create an unpredictable and confusing user experience across different platforms.

\textbf{Broader implications for user privacy and regulatory compliance.} The observed uneven implementation of privacy control across websites leads to an inconsistent and often subpar user experience, challenging the notion of comprehensive data protection. This study, by taking a holistic approach by examining the usability of four distinct types of privacy controls across multiple visit scenarios and website categories, highlights the systemic nature of these usability challenges, rather than isolated incidents. These pervasive flaws directly conflict with regulatory mandates like GDPR, which impose strict requirements on services and platforms, including mandating explicit user consent for data processing, and the broader global emphasis on data protection. When controls are difficult to find, understand, or offer limited choices, the spirit of user consent and control is undermined, despite the presence of privacy-related features.

\section{Conclusion}
\label{sec:conclusion}
Our investigation into 18,628 websites reveals widespread usability challenges in web privacy controls that directly impede users' ability to manage their data effectively. We found that privacy controls are frequently difficult to discover and cumbersome to access, often requiring excessive clicks, thus hindering user awareness and increasing the effort required for privacy management. A critical lack of multilingual support and navigational aids (e.g., tables of contents) in privacy policies and settings prevents users from fully understanding complex privacy information, leading to uninformed decisions. Moreover, many websites offer limited and non-meaningful choices, sometimes restricting users to `Accept' only or providing inadequate guidance on managing settings, thereby curtailing user autonomy and control over their data.

These findings underscore a persistent gap between current privacy control implementations and user expectations for transparent, accessible, and user-friendly data management tools, highlighting the urgent need for standardized and human-centric privacy practices across the web to align with evolving data protection regulations like GDPR. This study lays the foundation for future research into developing privacy tools that empower users, providing them with meaningful choices and greater control over their data.

\balance
\bibliographystyle{ACM-Reference-Format}
\bibliography{Main}
\newpage

\appendices
\section*{Appendix}\label{sec:appendix}
\setcounter{table}{0}
\renewcommand{\thetable}{A\arabic{table}}


\begin{table*}[!b]
  \centering
  \caption{Guest visit scenario and web crawl details.}
  \label{tab:webcrawltemplateguest}
  \small
  \vspace{-0.4cm}
  \resizebox{\textwidth}{!}{
    \begin{tabular}{p{0.8cm}|p{8cm}|p{6cm}}
      \toprule
      \textbf{Steps} & \textbf{Web Crawl Action} & \textbf{Web Crawl Metrics} \\
      \hline
      1 & Initiate a web crawl by entering the domain name in a search engine. & - \\
      \hline
      \multicolumn{3}{c}{\textbf{Privacy Nudges}} \\
      \hline
      1.1 & Identify the presence of privacy nudges during the crawl. & - \\
      \hline
      & If privacy nudges are present, extract information using the crawler. & - \\
      \hline
      1.1.1 & Determine the window screen location of privacy nudges. & Awareness \\
      1.1.2 & Identify types of triggers for privacy nudges. & Awareness \\
      1.1.3 & Extract information on the types of privacy nudges and their content. & Functionality \\
      1.1.4 & Collect details on privacy actions prompted by nudges. & Choice \\
      1.1.5 & Check for detailed explanations accompanying privacy actions. & Choice \\
      \hline
      \multicolumn{3}{c}{\textbf{Privacy Notices}} \\
      \hline
      1.2 & Identify the presence of privacy notices during the crawl. & - \\
      \hline
      & If privacy notices are present, extract information using the crawler. & - \\
      \hline
      1.2.1 & Determine the window screen location of privacy notices. & Awareness \\
      1.2.2 & Identify types of triggers for privacy notices. & Awareness \\
      1.2.3 & Extract information on the types of privacy information covered by notices. & Functionality \\
      \hline
      \multicolumn{3}{c}{\textbf{Privacy Policies}} \\
      \hline
      1.3 & Identify the presence of a privacy policy before login. & - \\
      \hline
      & If a privacy policy is present, extract information using the crawler. & - \\
      \hline
      1.3.1 & Determine the window screen location of the privacy policy button. & Awareness \\
      1.3.2 & Assess the number of clicks needed to access the privacy policy. & Efficiency \\
      1.3.3 & Check if the privacy policy is available in various languages. & Comprehension \\
      1.3.4 & Assess if the privacy policy contains contents to guide reading. & Comprehension \\
      1.3.5 & Check for the presence of privacy setting options in the privacy policy. & Choice \\
      1.3.6 & Identify links to privacy settings within the privacy policy. & Choice \\
      1.3.7 & Assess the clarity of privacy setting guidance in the privacy policy. & Choice \\
      1.3.8 & Determine the regulations followed by the privacy policy. & - \\
      \hline
      \multicolumn{3}{c}{\textbf{Privacy Settings}} \\
      \hline
      1.4 & Identify the presence of a link to privacy settings before login. & - \\
      \hline
      & If a link to privacy settings is present, extract information using the crawler. & - \\
      \hline
      1.4.1 & Determine the window screen location of the privacy setting button. & Awareness \\
      1.4.2 & Assess the number of clicks needed to access privacy setting options. & Efficiency \\
      1.4.3 & Check if privacy settings are available in various languages. & Comprehension \\
      1.4.4 & Evaluate the aspects of privacy settings covered. & Functionality \\
      \bottomrule
    \end{tabular}
  }
\end{table*}

\begin{table*}[htbp]
  \centering
  \caption{Sign-up user visit scenario and web crawl details.}
  \label{tab:webcrawltemplatesignup}
  \small
  \vspace{-0.4cm}
  \resizebox{\textwidth}{!}{
    \begin{tabular}{p{0.8cm}|p{8cm}|p{6cm}}
      \toprule
      \textbf{Steps} & \textbf{Web Crawl Action} & \textbf{Web Crawl Metrics} \\
      \hline
      2 & Identify if there is an option on the website to create a user account. If yes, create a user account with an alias and email address. & - \\
      \hline
      \multicolumn{3}{c}{\textbf{Privacy Nudges}} \\
      \hline
      2.1 & Identify the presence of privacy nudges during the crawl. & - \\
      \hline
      & If privacy nudges are present, extract information using the crawler. & - \\
      \hline
      2.1.1 & Determine the window screen location of privacy nudges. & Awareness \\
      2.1.2 & Identify types of triggers for privacy nudges. & Awareness \\
      2.1.3 & Extract information on the types of privacy nudges and their content. & Functionality \\
      2.1.4 & Collect details on privacy actions prompted by nudges. & Choice \\
      2.1.5 & Check for detailed explanations accompanying privacy actions. & Choice \\
      \hline
      \multicolumn{3}{c}{\textbf{Privacy Notices}} \\
      \hline
      2.2 & Identify the presence of privacy notices during the crawl. & - \\
      \hline
      & If privacy notices are present, extract information using the crawler. & - \\
      \hline
      2.2.1 & Determine the window screen location of privacy notices. & Awareness \\
      2.2.2 & Identify types of triggers for privacy notices. & Awareness \\
      2.2.3 & Extract information on the types of privacy information covered by notices. & Functionality \\
      \hline
      \multicolumn{3}{c}{\textbf{Privacy Policies}} \\
      \hline
      2.3 & Identify the presence of a privacy policy during account creation. & - \\
      \hline
      & If a privacy policy is present, extract information using the crawler. & - \\
      \hline
      2.3.1 & Determine the window screen location of the privacy policy button/link. & Awareness \\
      2.3.2 & Assess the number of clicks needed to access the privacy policy. & Efficiency \\
      2.3.3 & Check if the privacy policy is available in various languages. & Comprehension \\
      2.3.4 & Assess if the privacy policy contains contents to guide reading. & Comprehension \\
      2.3.5 & Check for the presence of privacy setting options in the privacy policy. & Choice \\
      2.3.6 & Identify links to privacy settings within the privacy policy. & Choice \\
      2.3.7 & Assess the clarity of privacy setting guidance in the privacy policy. & Choice \\
      2.3.8 & Determine the regulations followed by the privacy policy. & - \\
      \hline
      \multicolumn{3}{c}{\textbf{Privacy Settings}} \\
      \hline
      2.4 & Identify the presence of a link to privacy settings during account creation. & - \\
      \hline
      & If a link to privacy settings is present, extract information using the crawler. & - \\
      \hline
      2.4.1 & Determine the window screen location of the privacy setting button/link. & Awareness \\
      2.4.2 & Assess the number of clicks needed to access privacy setting options. & Efficiency \\
      2.4.3 & Check if privacy settings are available in various languages. & Comprehension \\
      2.4.4 & Evaluate the aspects of privacy settings covered. & Functionality \\
      \bottomrule
    \end{tabular}
  }
\end{table*}

\begin{table*}[htbp]
  \centering
  \caption{Registered user visit scenario and web crawl details.}
  \label{tab:webcrawltemplateregistered}
  \small
  \vspace{-0.4cm}
  \resizebox{\textwidth}{!}{
    \begin{tabular}{p{0.8cm}|p{9.0cm}|p{2.3cm}}
      \toprule
      \textbf{Steps} & \textbf{Web Crawl Action} & \textbf{Web Crawl Metrics} \\
      \hline
      3 & Visit the website with log-in status. & - \\
      \hline
      \multicolumn{3}{c}{\textbf{Privacy Nudges}} \\
      \hline
      3.1 & Identify the presence of privacy nudges during the crawl. & - \\
      \hline
      & If privacy nudges are present, extract information using the crawler. & - \\
      \hline
      3.1.1 & Determine the window screen location of privacy nudges. & Awareness \\
      3.1.2 & Identify types of triggers for privacy nudges. & Awareness \\
      3.1.3 & Extract information on the types of privacy nudges and their content. & Functionality \\
      3.1.4 & Collect details on privacy actions prompted by nudges. & Choice \\
      3.1.5 & Check for detailed explanations accompanying privacy actions. & Choice \\
      \hline
      \multicolumn{3}{c}{\textbf{Privacy Notices}} \\
      \hline
      3.2 & Identify the presence of privacy notices during the crawl. & - \\
      \hline
      & If privacy notices are present, extract information using the crawler. & - \\
      \hline
      3.2.1 & Determine the window screen location of privacy notices. & Awareness \\
      3.2.2 & Identify types of triggers for privacy notices. & Awareness \\
      3.2.3 & Extract information on the types of privacy information covered by notices. & Functionality \\
      \hline
      \multicolumn{3}{c}{\textbf{Privacy Settings}} \\
      \hline
      3.3 & Identify the presence of a privacy setting link/button after login. & - \\
      \hline
      & If a link/button to privacy settings is present, extract information using the crawler. & - \\
      \hline
      3.3.1 & Determine the window screen location of the privacy setting button/link. & Awareness \\
      3.3.2 & Assess the number of clicks needed to access privacy setting options. & Efficiency \\
      3.3.3 & Check if privacy settings are available in various languages. & Comprehension \\
      3.3.4 & Evaluate the aspects of privacy settings covered. & Functionality \\
      \bottomrule
    \end{tabular}
  }
\end{table*}

\begin{table*}[htbp]
    \centering
    \caption{Detailed category mapping with website counts.}
    \label{tab:category-mapping}
    \small
    \vspace{-0.4cm}
    \begin{tabular}{l|c|l}
        \hline
        \textbf{Original Category} & \textbf{Number of Websites} & \textbf{Mapped Category} \\
        \hline
        Information Technology & 3727 & Technology \\
        Search Engines and Portals & 407 & Technology \\
        File Sharing and Storage & 318 & Technology \\
        Webmail & 37 & Technology \\
        \hline
        Business & 3444 & Business and Finance \\
        Finance and Banking & 895 & Business and Finance \\
        Real Estate & 93 & Business and Finance \\
        \hline
        Entertainment & 997 & Entertainment \\
        Arts and Culture & 396 & Entertainment \\
        Games & 584 & Entertainment \\
        Internet Radio and TV & 138 & Entertainment \\
        Gambling & 94 & Entertainment \\
        Streaming Media and Download & 85 & Entertainment \\
        \hline
        News and Media & 1538 & News and Social Media \\
        Newsgroups and Message Boards & 248 & News and Social Media \\
        Personal Websites and Blogs & 175 & News and Social Media \\
        Social Networks & 98 & News and Social Media \\
        Chats \& Messengers & 45 & News and Social Media \\
        Dating & 34 & News and Social Media \\
        \hline
        Pornography & 1907 & Adult Content \\
        \hline
        Shopping & 1432 & Shopping \\
        \hline
        Education & 1274 & Education \\
        \hline
        Travel & 310 & Lifestyle and Sports \\
        Sports & 309 & Lifestyle and Sports \\
        Health and Wellness & 198 & Lifestyle and Sports \\
        Automotive & 175 & Lifestyle and Sports \\
        \hline
        Government and Legal Organizations & 742 & Government \\
        \hline
        Advertising & 181 & Other \\
        Meaningless Content & 143 & Other \\
        Newly Observed Domain & 143 & Other \\
        Not Rated & 73 & Other \\
        Potentially Unwanted Program & 63 & Other \\
        Religious & 50 & Other \\
        \hline
        Illegal or Unethical & 160 & Malicious Content \\
        Malicious Websites & 35 & Malicious Content \\
        Phishing & 33 & Malicious Content \\
        \hline
    \end{tabular}
\end{table*}

\end{document}